\documentclass[twocolumn,showpacs,preprintnumbers,amsmath,amssymb]{revtex4}
\usepackage{graphicx}
\usepackage{dcolumn}
\usepackage{bm}
\usepackage{mathrsfs}

\begin{document}
\preprint{APS/123-QED}

\title{Co-existing structures in $^{105}$Ru}

\author{S.~Lalkovski$^1$\footnote{email: stl@phys.uni-sofia.bg}, D.~Ivanova$^1$, E.~A.~Stefanova$^2$,
A.~Korichi$^3$,  P.~Petkov$^2$, J.~Kownacki$^4$, T.~Kutsarova$^2$, A.~Minkova$^1$, D.~Bazzacco$^5$, 
M.~Bergstr\"om$^6$, A.~G\"orgen$^7$\footnote{Present address: DAPNIA/SPhN, CEA-Saclay, Gif-sur-Yvette, France}, 
B.~Herskind$^6$, H.~H\"ubel$^7$, A.~Jansen$^7$, S.~Kisyov$^1$, T.~L.~Khoo$^8$, F.~G.~Kondev$^9$, 
A.~Lopez-Martens$^2$,  
Zs.~Podoly\'ak$^{10}$\footnote{Present address: Department of Physics, University of Surrey, Guildford GU27XH, UK},
G.~Sch\"onwasser$^7$, O.~Yordanov$^2$}

\affiliation{
$^1$Faculty of Physics, University of Sofia, Sofia 1164, Bulgaria\\
$^2$Institute for Nuclear Research and Nuclear Energy, Bulgarian Academy of Science, Sofia 1784, Bulgaria\\
$^3$CSNSM Orsay, IN2P3/CNRS, F-91405, France\\
$^4$Heavy-Ion Laboratory, University of Warsaw, ul. "Pasteura" 5A, 02-093 Warzsawa, Poland\\
$^5$INFN, Sezione di Padova, I-35131 Padova, Italy\\
$^6$The Niels Bohr Institut, Blegdamsvej 17, DK-2100 Copenhagen, Denmark\\
$^7$Helmholtz-Institut f\"ur Strahlen -und Kernphysik, Universit\"at Bonn, Nussallee 14-16, D-53115 Bonn, Germany\\
$^8$Physics Division, Argonne National Laboratory, Argonne, Illinois 60439, USA\\
$^9$Nuclear Engineering Division, Argonne National Laboratory, Argonne, Illinois 60439, USA\\
$^{10}$INFN, Laboratori Nazionali di Legnaro, I-35020 Padova, Italy\\
}

\date{\today}

\begin{abstract}
New positive-parity states, having a band-like structure, were observed in 
$^{105}$Ru. The nucleus was produced in induced fission reaction and the prompt 
$\gamma$-rays, emitted from the fragments, were detected by the EUROBALL III 
multi-detector array. The partial scheme of excited $^{105}$Ru levels is analyzed
within the Triaxial-Rotor-plus-Particle approach.
\end{abstract}

\maketitle
\section{Introduction}
$^{105}$Ru is located on the Segr\'e chart between its heaviest stable isotope 
$^{104}$Ru \cite{Bl07} and the most exotic $^{117,118,119}$Ru nuclei, produced
in relativistic fission \cite{La12, Ka12, PA13}. Being just on the edge of the 
line of $\beta$-stability, only few experimental methods can be used to populate 
its excited states. So far, the nucleus was studied in the $^{105}$Tc $\beta$-decay 
\cite{Su75}, $^{104}$Ru(d,p) reaction \cite{MK76, Fo71} and $n$-capture on $^{104}$Ru 
\cite{Gu78, Hr74}. However, these reaction mechanisms are highly selective and 
populate only low-spin states. In the 1990s the high-resolution high-granularity 
multidetector $\gamma$-ray arrays become available, which have enabled the use 
of induced fission reactions for $\gamma$-ray spectroscopy, providing the
opportunity to fill in the gap of transitional nuclei situated between the 
line of beta stability and the most exotic neutron-rich nuclei produced in 
fission. By using induced fission reaction, the intruder negative-parity band in 
$^{105}$Ru was observed for the first time and extended to $(31/2^-)$ \cite{Fo98}. 
The present work reports on new results for $^{105}$Ru, obtained also from induced 
fission. Two positive-parity bands were observed on top of the known $7/2_1^+$ 
and $5/2_2^+$ states, which help to parametrize the Rigid-Triaxial-Rotor-plus-Particle
model and test its applicability to the low-lying low-spin states observed  prior 
to our study in $^{105}$Ru.

\begin{figure*}[ht]
\rotatebox{0}{\scalebox{0.45}[0.45]{\includegraphics{Ru105_specs.eps}}}
\hspace{0.75cm}
\rotatebox{0}{\scalebox{0.28}[0.28]{\includegraphics{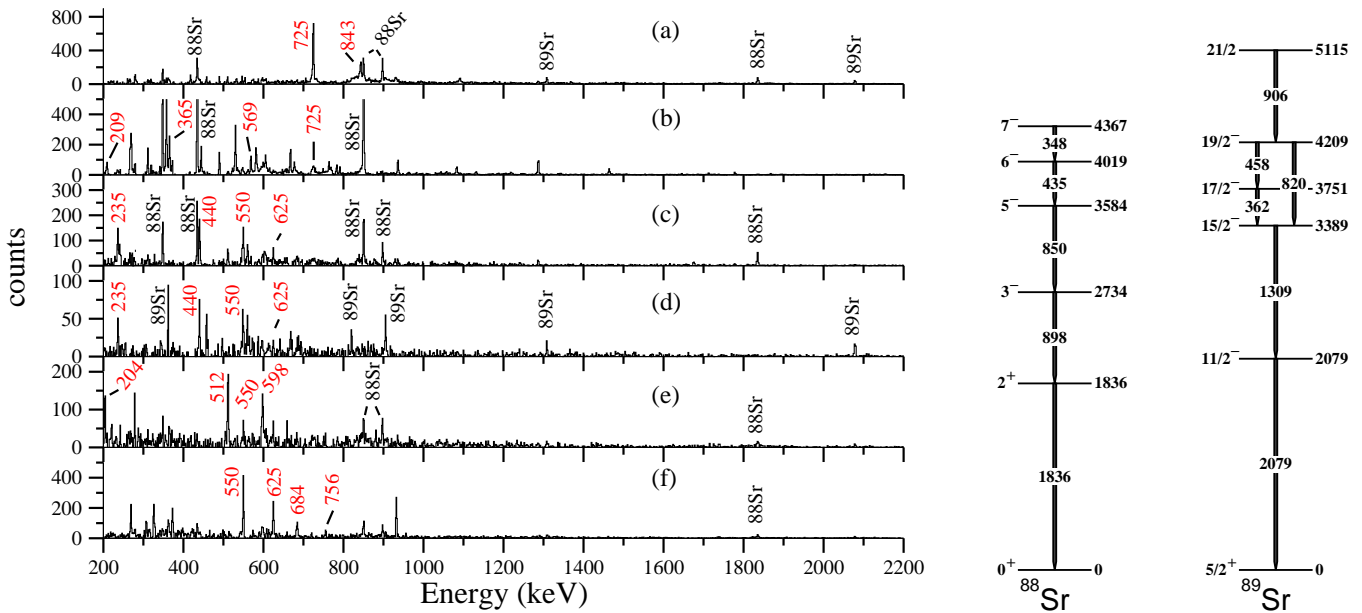}}}
\caption[]{(color on-line) {\bf Left hand side:} $\gamma$-ray spectra, gated on: 
(a) 365 and 569-keV transitions; (b) 1836 and 898-keV transitions; (c) 209 and 
1836 or 209 and 898-keV transitions; (d) 209 and 2079-keV or 209 and 1309-keV 
transitions; (e) 209 and 235-keV transitions; (f) 209 and 440-keV transitions. 
{\bf Right hand side:} Partial level schemes of $^{88}$Sr and $^{89}$Sr according to 
Refs.~\cite{St00} and \cite{St01}.}
\label{specs}
\end{figure*}

\section{Experiment and Data analysis}
$^{105}$Ru was produced as a fission fragment in the disintegration of the 
$^{198}_{\ 82}$Pb compound 
nucleus, which was synthesized in the $^{30}_{14}$Si+$^{168}_{\ 68}$Er reaction 
at a beam energy of $E(^{30}$Si)=142 MeV. In order to stop the recoils, the 1.15 
mg/cm$^2$ thick $^{168}$Er target was deposited on a 9 mg/cm$^2$ gold backing. 
The $\gamma$-rays, emitted by the fission products, were detected by the EUROBALL III 
multidetector array comprising 30 single HPGe detectors, 26 Clover and 15 Cluster 
detectors with anti-Compton shields. The acquisition system was triggered by triple 
$\gamma - \gamma - \gamma$ coincidences. 3D cubes were sorted and analyzed with
the RADWARE software \cite{Ra95}. Extended level scheme of $^{194}_{\ 82}$Pb, 
produced in the $4n$ fusion-evaporation channel, was previously reported in 
Ref.~\cite{Ku07}. Data for $^{98,100,102}_{\quad \quad \ \ 42}$Mo and   
$^{109,111}_{\quad \ \ 46}$Pd, produced in the same experiment as fission fragments, 
are published in Refs.~\cite{La07} and \cite{St12}. Their respective complementary 
$_{40}$Zr and $_{36}$Kr fragments were found in the 4 -- 6$n$ fusion-fission channels.

\begin{figure}[ht]
\rotatebox{0}{\scalebox{0.4}[0.4]{\includegraphics{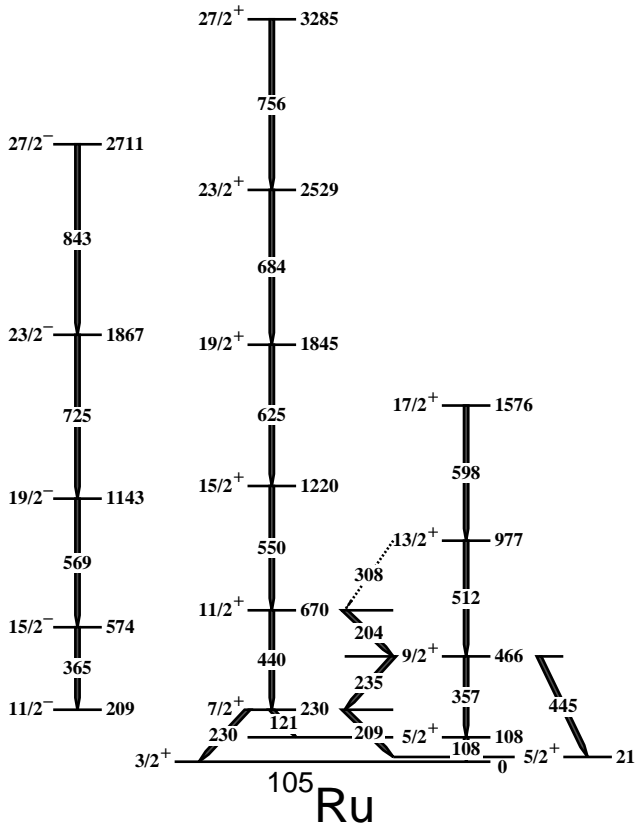}}}
\caption[]{Partial level scheme of $^{105}$Ru as observed in the present study. 
Level and $\gamma$-ray energies are rounded of values from Table~\ref{105Pd_tabl}. 
Spin and parity assignments to the levels, observed in the present work, are only 
tentative. Arguments for these assignments are given in the text.}
\label{105Pd_lsc}
\end{figure}

Sample energy spectra, obtained from the present experiment, are shown in 
Fig.~\ref{specs} and the partial level scheme of $^{105}$Ru, based on the 
coincidence measurements is present in Fig.~\ref{105Pd_lsc}.

In the present experiment, the complementary fragments to $^{105}_{\ 44}$Ru 
are $_{38}$Sr nuclei, given that no proton evaporation can
occur in the induced fission reactions. In order to identify its most probable 
complementary Sr isotopes prompt transitions with energies of 365, 569, 725 and 
843 keV from the negative-parity band in $^{105}$Ru \cite{Fo98} were used. 
Fig.~\ref{specs}(a) shows a sample $\gamma$-ray spectrum in coincidence with the
365 and 569-keV transitions. The 2079 and 1309-keV transitions in $^{89}$Sr \cite{St01}, 
and the 1836 and 898-keV transitions in $^{88}$Sr \cite{St00}, 
which correspond to the 4$n$ and $5n$ fission-fission channels, respectively, 
were observed in coincidence with the $^{105}$Ru negative-parity band members. 
The $^{88,89}$Sr being complementary fragments to $^{105}$Ru is 
consistent with the number of neutrons, evaporated prior to the $\gamma$-ray 
emission, observed in the cases of $^{109,111}$Pd \cite{St12} and $^{98,100,102}$Mo 
fragments \cite{La07}. The partial level schemes of $^{88,89}$Sr are shown in 
Fig.~\ref{specs}.

To search for the positive-parity yrast states in $^{105}$Ru, coincidence spectra 
gated on the 1309 and 2079-keV transitions in $^{89}$Sr and 898 and 1836-keV in
 $^{88}$Sr were simultaneously studied. A sample spectrum, gated on the 1836 and
898-keV transitions in $^{88}$Sr, is shown in Fig.~\ref{specs}(b). A weak 209-keV 
transition was observed both in the $^{88}$Sr and $^{89}$Sr gated spectra, which 
suggests that it is in a complementary Ru nucleus. Indeed, this transition 
de-excites the $7/2_1^+$ state in $^{105}$Ru, produced in ($n,\gamma$) \cite{Gu78, Hr74} 
and $\beta$-decay \cite{Su75}. There, the 209-keV transition is the strongest 
decay branch from the 230-keV $7/2_1^+$ level. Further, cross-coincidence gates, 
imposed on the 209-keV transition and transitions in the complementary $^{88}$Sr 
and $^{89}$Sr nuclei, reveal that the 209-keV transition is in coincidence with 
transitions of 440, and 550 keV. Sample spectra are shown in Fig.~\ref{specs}(c,d). 
Spectra, gated on the 209 and 440 or 550-keV transitions show they are part of 
the band extended up to 3285-keV in  Fig.~\ref{105Pd_lsc}. In the present data, 
the 230-keV level decays also by a second branch of weak transitions with energies 
of 121 and 108-keV, which were also observed in the $(n,\gamma)$ \cite{Hr74, Gu78} 
and $\beta$-decay \cite{Su75} data, confirming that the band is a part of the 
$^{105}$Ru level scheme. 

Two more transitions with energies of 204 and 235 keV were observed in coincidence 
with the 550 and 625-keV $\gamma$ rays. The last two spectra in Fig.~\ref{specs}(e) 
and (f) show that the 209 and 235-keV transitions link the band based on the 
$7/2^+$ state to a second sequence of transitions on top of the second $5/2^+$ 
state at 108 keV. 

The $\gamma$-ray energies ($E_\gamma$) and their relative intensities ($I_\gamma$), 
observed in the present study, are listed in Table~\ref{105Pd_tabl} along with the 
$\gamma$-ray energies ($E_\gamma^{ENS}$) and ($I_\gamma^{ENS}$), adopted in 
Ref.~\cite{FJ05}. The intensities $I_\gamma$ are normalized with respect to the 
intensity of the 550.1-keV transition, while the $I_\gamma^{ENS}$, from the 
adopted in ENSDF gammas, are normalized with respect to the intensity of the 
strongest decay branch to each level. Due to the poor statistics, no $I_\gamma$
were obtained for the 108.4, 121.1, and 229.8-keV transitions. For the purpose of
the discussion below, the branching ratios from the adopted gammas will be used
in these cases.

The spin and parity ($J^\pi$) assignments to the states, below the 230-keV level, 
are based on the $\beta$-decay feeding \cite{Su75}, on the decay branches of 
primary transitions observed from the $n$-capture state at 5.9 MeV 
\cite{Hr74, Gu78}, the  $L$ transfer in $(d,p)$ reaction \cite{Fo71, MK76} and 
the $\gamma$-decay pattern to states with known spins and parities. 
The $J^\pi$ assignments to the higher-energy states are 
based on the observed band structure, and for each level -- on the $\gamma$-decay 
branches to states with known spin and parity assignments. The $J^\pi$ assignments 
in the present work are also supported by the systematics of the positive-parity 
states in the Ru isotopic chain. Only $M1$ and $E2$ nature is assumed for the 
prompt $\gamma$-rays above the $7/2_1^+$ state. Given that the induced fission
reaction populates mainly yrast states, it can be expected that at moderate and 
high energies there will be a little overlap between the states produced in the 
present experiment and in the $\beta$-decay or $n$-capture primary transitions.

\begin{table}[ht]
\caption{\label{105Pd_tabl}Gamma-decay properties: initial level energy $E_i$ 
in (keV), obtained from a least squares fit to the $\gamma$-ray energies $E_\gamma$,
observed in the present work; spin/parity assignments $J^\pi_i$ to the levels in $^{105}$Ru; 
$\gamma$-ray energies $E_\gamma$ in (keV) and relative intensities $I_\gamma$, 
normalized with respect to the intensity of the 550-keV transition. The 
$J^{\pi,ENS}$ assignments, $\gamma$-ray energies $E_\gamma^{ENS}$ and intensities 
$I_\gamma ^{ENS}$, adopted by NNDC \cite{FJ05}, are also listed for completeness. 
The level energies, adopted by NNDC \cite{FJ05}, deviate by less then one keV from 
$E_i$ and hence not given in a separate column.}
\begin{tabular}{llllllllll}
\hline\hline
$E_i$ & $J_i^\pi$    &   $E_\gamma$& $I_\gamma$  &   $J^{\pi, ENS}$ & $E_\gamma^{ENS}$& $I_\gamma ^{ENS}$ \\
\hline
20.9(11)    &  $5/2^+$&        &         & $5/2^+$ & 20.559 & 100\\
108.4(8)   &  $5/2^+$ &        &         & $5/2^+$ & 87.40  & $<1.1$ \\  
           &          & 108.4  &         &          & 107.945& 100 \\
229.8(8)   &  $7/2+$  & 121.2  &         & $7/2^+$ & 121.49 & 16(4)  \\
           &          & 208.6  &  76  (8)&          & 208.89 & 100 \\    
           &          & 229.8  &         &          & 229.51 & 16(4)\\ 
465.7(10)  &$(9/2+)$  & 235.4  & 68  (6) & $(3/2)^+$& 306.76 & 55(20)\\
           &          & 357.3  & 76 (8)  &           & 358.15 & 100\\
           &          &  445.1 &$\leq 12$&          & 445.81  & 71 (8)\\
           &          &        &         &          & 466.23  & 66(6)\\
670.1(11)  &$(11/2^+)$& 204.3  & 28  (3) & $\leq 5/2^+$& 397.8 & 8(4)  \\
           &          & 440.4  & 104 (9) &          & 562.7   & 100  \\
977.4(14)  &$(13/2^+)$& 511.7  & 42(5)   &          &         & \\
1220.2(15) &$(15/2^+)$& 550.1  & 100     &          &         & \\
1575.6(17) &$(17/2^+)$&  598.2 & $\leq$40&          &         &  \\
1845.1(18) &$(19/2^+)$&  624.9 & 68 (3)  &          &         &  \\
2529.3(21) &$(23/2^+)$&  684.2 & 29.6 (15)&         &         & \\
3285.4(23) &$(27/2^+)$&  756.1 & 13.8 (11)&         &         &  \\          
\hline\hline
\end{tabular}\\
\end{table}

By applying the above procedures, two of the $J^\pi$ assignments, made in the 
present work, differ significantly from the $J^{\pi,ENS}$ values adopted in 
Ref.~\cite{FJ05}. Thus, for example a 466-keV level is reported from $\beta$-decay, 
$^{104}$Ru$(d,p)$ and $^{104}$Ru($n,\gamma$) 
data. There, the level decays via a branch of four transitions with energies of 
307, 358, 446, and 466 keV to levels with $J^\pi=1/2^+$, $3/2^+$, and $5/2^+$. 
Also, the 466-keV level is fed by a primary transition from the $1/2^+$ $n$-capture 
state at 5.9 MeV which constrains the possible $J^\pi$ assignments to $J^\pi=1/2^+$ 
and $3/2^+$, with the last one being adopted in Ref.~\cite{FJ05}. 
In contrast, the 466-keV level, observed in the present study, decays via a 
strong 235-keV transition in addition to the main 357-keV decay branch and the 
307 and 466-keV decay branches were not observed at all. Therefore the state, 
observed in the present study, is tentatively interpreted as a $J^\pi=9/2^+$ 
state, different from the low-spin 466-keV level known prior our work. 

A similar situation occurs for the 670-keV level, which is observed 
to decay via 204 and 440-keV transitions in the present study, while in Ref.~\cite{FJ05} 
the level at 670 keV is de-excited by 398 and 563-keV transitions. Again, the 
low-spin assignment to the 670-keV level in NNDC is based on the feeding from 
the $n$-capture state at 5.9-MeV via a 5.24-MeV primary transition. In 
contrast to the NNDC data, the 670-keV level in the present study decays to the 
$7/2^+$ and $9/2^+$ states which, along with the assumption that it is an yrast 
state, leads to $J^\pi =11/2^+$ assignment to this level. 

Before closing this section, we should note that the intensity balance at the 
21-keV level, performed from the $n$-capture data \cite{Gu78}, leads to an 
experimental electron conversion coefficient $\alpha _{exp}\leq 25.2$, which along 
with the electron conversion coefficients calculated for a pure $M1$ or $E2$ 
20.56-keV transition $\alpha _{M1}=4.016$ and $\alpha _{E2}=409.4$ \cite{Ki08}, 
respectively, gives a multipole mixing ratio $\delta \leq 0.23$. The half-life 
$T_{1/2}=340$(15) ns of the first excited state was measured by the 
$143.25\gamma -20.55\gamma (t)$ delayed coincidences in the $^{105}$Tc
 $\beta$-decay \cite{Su75}, which leads to a hindered $M1$ component with 
$B(M1)= 2.7\times 10^{-4}$ W.u. and  possibly enhanced $E2$ component with 
$B(E2)\leq 30.7$ W.u.

The 164-keV level in Ref.~\cite{FJ05}, not observed in the present study, decays 
with $T_{1/2}=55 (7)$ ns via a week 55-keV transition to the 108-keV 
$J^\pi =5/2_2^+$ level as well as via a strong 143-keV $M1+E2$ transition to the 
21-keV $5/2_1^+$ level. In Ref.~\cite{FJ05}, $J^\pi =3/2^+, 5/2^+$ was assigned 
to this level. Hence, the reduced transition probabilities are 
$B(M1; 55\gamma)=1.01\times 10^{-4}$ (22) W.u. and 
$B(M1;143\gamma)=1.00 \times 10^{-4}$ (15) W.u. for the two decay branches, 
respectively. The respective $E2$ component of the 143-keV transition has
$B(E2;143\gamma)=0.27$ (13) W.u..
Thus, the two low-lying isomeric states, observed in $^{105}$Ru 
prior our study, decay via hindered $M1$ transitions.

\begin{figure*}[ht]
\rotatebox{90}{\scalebox{0.45}[0.45]{\includegraphics{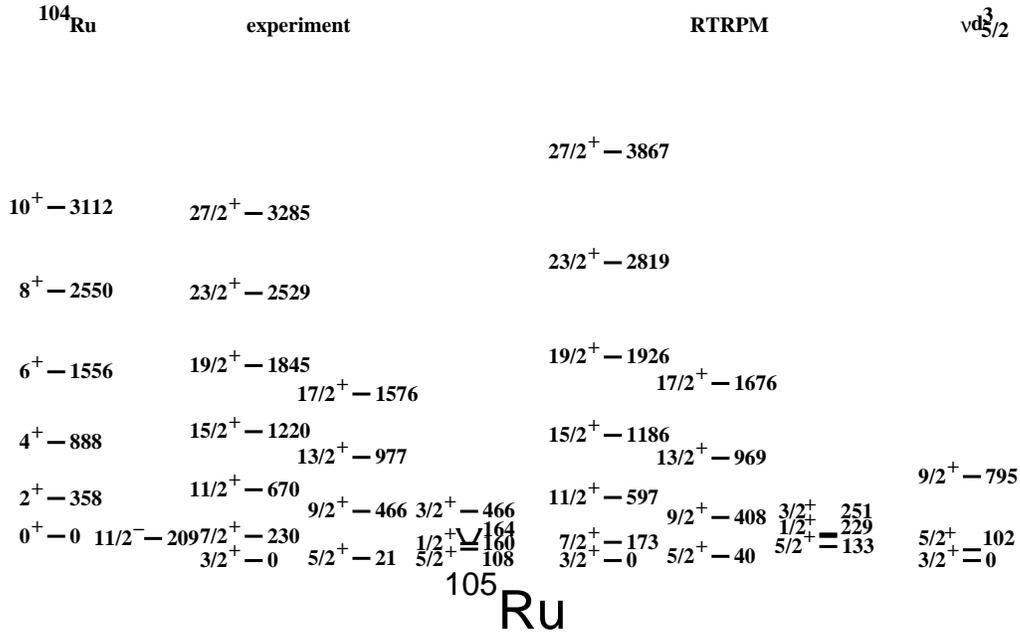}}}
\caption[]{\label{105th}Experimental and theoretical $^{105}$Ru level schemes. 
The $\epsilon _2= 0.24$, $\epsilon _4= -0.013$, $\gamma   = 20^\circ$, and 
$E_{2^+}  = 0.2$ MeV parameters were used to obtain
the Rigid Triaxial Rotor plus Particle Model (RTRPM) spectrum. The Empirical Shell 
model calculations within the $\nu d_{5/2}^3$ coupling scheme are parametrized 
with respect to the $^{104}$Ru data.}
\end{figure*}

\section{Discussion}
Even though the low-lying states in $^{105}$Ru are extensively studied via 
different experiments their structure is still not well understood. Thus, 
from (d,p) reactions \cite{MK76, Fo71}, large spectroscopic factors were obtained 
for the $J^\pi =5/2_1^+$ (21 keV), $1/2_1^+$ (160 keV), $11/2_1^-$ (209 keV), 
$7/2_1^+$ (230 keV) and $3/2_2^+$ (466 keV) states in $^{105}$Ru, shown in the 
experimental level scheme on Fig.~\ref{105th}, suggesting that they contain a large 
fraction of the single particle strength of the $\nu 2d_{5/2}$, $\nu 3s_{1/2}$, 
$\nu 1h_{11/2}$, $\nu 1g_{7/2}$, and $\nu 2d_{3/2}$ orbitals. However, $^{105}$Ru 
has a high level density at low energies \cite{FJ05} and the remaining 
single-particle strength is distributed over a larger number of states. In 
the shell model approach, some of these states can be interpreted as seniority 
$v = 3$ states \cite{Hy94}. Such an interpretation was already suggested in \cite{Hr74} 
for the ground state, in order to account for the small spectroscopic factor 
observed in the $(d,p)$ reaction. Indeed, $(\nu d_{5/2})^3$ calculations, shown
in Fig.~\ref{105th}, with a two-body matrix elements parametrized with respect 
to the neighboring $^{104}$Ru, reproduce correctly the $3/2^+$ ground state. The 
$5/2^+$ member of the multiplet is calculated at 102 keV above the ground state, 
which is consistent with the 108-keV state in the experimental level scheme and 
the small spectroscopic factor obtained from the $(d,p)$ reaction. Also, 
the empirical shell model calculations predict a $9/2^+$ level at 795 keV. 

Given that the $l$-forbidden $M1$ transitions in this mass region are typically 
hindered by two or three orders of magnitude \cite{An85}, the extra degrees of 
forbiddenness observed for the $M1$ transition from the 21-keV, $\nu d_{5/2}$ 
level is consistent with a more complex structure of the ground state, which could 
involve a $\nu d_{5/2}^3$ configuration. This scenario could be further tested 
if we knew the half-life of the 108-keV level, given that the transitions between 
the same multiplet members are hindered \cite{Ta63}. By closing this paragraph, 
it worth noting that the $\nu g_{7/2}$ orbit is also observed at low energy and 
the occurrence of the respective $j^3$ multiplet members would make the picture 
even more complicated.  Thus, to account for all single-particle orbits a 
detailed shell-model calculations are needed. 

An alternative approach to the problem would be to restrict the valence space as 
it is realized in the particle-core coupling models. In the weak-coupling model 
\cite{dS61}, discussed in the literature as a possible approach to the $^{105}$Ru
case, the excitations of an odd-mass nucleus is considered to be either 
single-particle or collective excitations of the even-even core. In this model, 
the $M1$ transitions between the same multiplet members  are forbidden while the 
$E2$ transitions are enhanced. This resembles the $^{105}$Ru case, however, 
the magnitude of the multiplets splitting in $^{105}$Ru is of the order of the 
first phonon energy which makes it difficult to identify the multiplet members. 
Also, the weak coupling model, which should work better for less deformed nuclei, 
fails in describing $^{101}$Ru \cite{Go73} suggesting that it might not be 
suitable for $^{105}$Ru too.

In the present work, the particle-core coupling concept will be further tested 
for $^{105}$Ru by using the Rigid-Triaxial-Rotor-plus-Particle model (RTRPM) 
\cite{LLR78}. This model seems to be appropriate for the case of $^{105}$Ru, 
given that the nucleus is located in an island of triaxal nuclei.

\begin{figure}[ht]
\rotatebox{-90}{\scalebox{0.3}[0.3]{\includegraphics{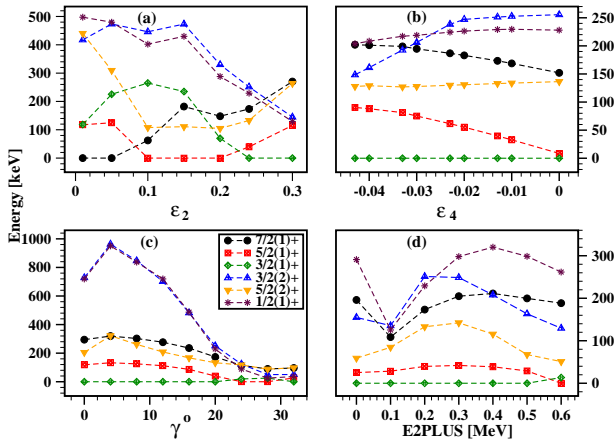}}}
\caption[]{(color on-line) Evolution of low-lying states in $^{105}$Ru as a function of 
the Rigid-Triaxial-Rotor-plus-Particle model parameters (a) $\epsilon _2$; 
(b) $\epsilon _4$; (c) $\gamma$; and (d) $E_{2^+}$. In each subfigure, the 
unfitted parameters were held fixed to the values given in Fig.~\ref{105th}.
}
\label{RTRPMC}
\end{figure}

\subsection{Rigid Triaxial Rotor plus Particle model calculations}

Theoretical calculations for $^{105}$Ru were performed with the RTRPM in a strong 
coupling basis  \cite{LLR78}. The single-particle wave functions were calculated 
with GAMPN code, which is part of the ASYRMO package \cite{SR92}. A Standard 
set of the Nilsson parameters \cite{BR85} $\kappa _4= 0.070$, $\mu _4 = 0.39$ 
and $\kappa _5= 0.062$ and $\mu _5 = 0.43$ was used. The level energies were 
calculated with ASYRMO \cite{SR92}, which diagonalyzes the particle+triaxial 
rotor Hamiltionian. The quadrupole 
deformation $\epsilon _2$  and the moment-of-inertia $\hbar^2 / 2\Im = E_{2^+}/6$ 
parameters were deduced from the neighboring even-even nuclei. A Corriolis 
attenuation factor $\xi = 0.7$ was also used to obtain a better description of the
band structure. The pairing was parametrized via $GN0=22.0$, $GN1=8.0$ and 
$IPAIR=5.0$. In order to obtain a better fit to the experimental data, 
$\epsilon _2$, $\epsilon _4$, $\gamma$ and $E_{2^+}$ parameters were varied.  A 
good fit to the experimental level energies was obtained with $\epsilon _2= 0.24$, 
$\epsilon _4= -0.013$, $\gamma   = 20^\circ$, and $E_{2^+}  = 0.2$ MeV. This set 
of parameters is consistent with the respective $\epsilon _2$ and $\gamma$, 
obtained from the neighboring even-even nuclei. A comparison of the experimental 
and theoretical RTRPM level energies is shown in Fig.~\ref{105th} and an example 
of the variation procedures applied for $^{105}$Ru is present in Fig.~\ref{RTRPMC}.

The level energy dependence on $\epsilon _2$ is shown in Fig.~\ref{RTRPMC}(a).
The figure shows also that $3/2^+$ is the ground state in $^{105}$Ru only at 
large deformations, i.e. $\epsilon \geq 0.24$. Also, the relative position of 
the low-lying states strongly depends on the deformation parameter $\epsilon _2$.

Figure \ref{RTRPMC}(b) shows that $3/2^+$ is the ground state of $^{105}$Ru in a wide 
range of $\epsilon _4$. This parameter slightly affects also the behavior of the 
$1/2_1^+$ and $5/2_2^+$ levels, while its influence on the $3/2_2^+$,
$5/2_1^+$ and $7/2_1^+$ level energies is stronger.

Figure~\ref{RTRPMC}(c) shows that the $^{105}$Ru level energies strongly depend 
on the parameter of triaxial deformation $\gamma$ for $5^\circ \leq \gamma \leq 25^\circ$. 
This is well pronounced for the $1/2_1^+$ and $5/2_2^+$ levels and to a lower 
extent for the $7/2_1^+$, $5/2_1^+$ and $3/2_1^+$ levels. Except for 
$\gamma =24^\circ$ to 26$^\circ$, $3/2^+$ is the ground state in the entire 
range of $0^\circ \leq \gamma \leq 30^\circ$ . 

Figure~\ref{RTRPMC}(d) shows that the ground state is less sensitive to the moment 
of inertia parameter and $3/2^+$ is the ground state for $E_{2^+} <$0.6 MeV. 
Depending on the effect of the moment-of-inertia parameter on the level energies, 
two subset of states can be distinguished. The first group of levels is formed by 
the $3/2_1^+$ and $5/2_1^+$ levels, which are almost independent on this parameter. 
The second sub-set is formed by the $1/2^+$ and the $7/2^+$ states with energies 
strongly dependent on the moment-of-inertia parameter. $3/2^+_2$ and $5/2^+_2$ 
have a more intermediate trend with respect to the $E_{2^+}$ parameter.

As shown in Fig.~\ref{105th}, a good overall description of the experimental 
bands, based on the $7/2^+$ and $5/2_1^+$ excited states, is achieved up to 
$19/2^+$. 
At higher spins, the experimental bands are more squeezed than the theoretical. 
This effect could be explained by the backbending usually observed in the 
positive parity bands of the odd-$N$, even-$Z$ nuclei in this mass region. 
Indeed, the positive-parity sequence, based on the $7/2^+$ state, closely 
resembles the yrast band in $^{104}$Ru, as shown in Fig.~\ref{105th}, where 
backbending is observed.

The energy of the non-yrast $5/2_2^+$ and $1/2^+$ states, experimentally observed 
close to the ground state, is also well reproduced. The only major discrepancy 
between the theory and the experiment at low energies, is in the $3/2_2^+$ level 
energy which is underestimated by the calculations by approximately 200 keV. This 
is, somewhat surprising, given this level is expected to be of single-particle 
nature and hence should be in the model space.

The $M1$ and $E2$ transition probabilities were calculated with PROBAMO \cite{SR92}. 
The standard value of $GSFAC=0.60$ for the modification of the free gyromagnetic 
factor was used. The magnetic moment of the $3/2^+$ ground state, obtained from 
the RTRPM calculations is $\mu = -0.13 \mu_N$, which is consistent with the 
experimental value $\mu =(-)0.32(+8-20) \mu_N$ \cite{Ha81}. 

The $B(E2)=20$ W.u., calculated for the $5/2_1^+\rightarrow 3/2^+_1$ isomeric 
transition, is consistent with the experimental value 30.7 W.u., and shows that
the $5/2_1^+$ wave function has a collective component. However, the 
$B(M1)=0.010$ W.u., calculated with RTRPM, is highly overestimated given that the 
experimental $B(M1)= 3\times 10^{-4}$ W.u. Even though, the theoretical $B(M1)$ 
is enhanced with respect to the experimental value, this is still consistent with 
the experimental data. In the case where the initial state is a $\nu d_{5/2}$ 
state and the ground state involves a $\nu d_{5/2}^3$ component, which is outside 
the RTRPM model space, extra degrees of forbiddenness can be expected.

The 55-ns isomer, observed at 164-keV in $^{105}$Ru \cite{FJ05} has even a more
obscure structure. This is partially because the existing experimental data 
does not allow a specific $J^\pi$ assignment to that level \cite{FJ05} and also 
the $T_{1/2}=55$ ns was assigned rather to a $1/2^+$ level \cite{HLB78} than to 
the 164-keV level, assuming the 143-keV transition is a $l$-forbidden transition. 
Indeed, the RTRPM calculations does predict a $3/2^+$ state at 251 keV, which 
could be the 164-keV state in Ref.~\cite{FJ05}, however, it decays to the first 
and the second $5/2^+$ states via transitions with $B(M1)=2\times 10^{-3}$ W.u. 
and 3.3$\times 10^{-2}$ W.u., 
respectively. The extra degree of hindrance, observed in the experimental $B(M1)$ 
to the $5/2_2^+$ state, could be related to the structure of the final state 
given it is a member of the $\nu d_{5/2}^3$  multiplet. Similarly, the
$B(M1)=0.03\times 10^{-2}$ W.u. decay branch to the ground state, calculated with
the RTRPM, is not experimentally observed. Hence, to completely understand 
the structure of the 164-keV state more experimental data is needed, including 
unambiguous data for the $J^\pi$ assignments and a thorough study of its decay 
branches.

\section{Conclusions}
$^{105}$Ru was produced in induced fission reaction. Its level scheme was 
extended up to $27/2^+$ and a new positive-parity band was identified. 
Rigid-Triaxial-Rotor-plus-particle model calculations were performed for 
$^{105}$Ru. The model was parametrized to fit the level energies, known from 
literature, as well as the data obtained in the present study. In the medium-spin 
regime, i.e. for $J^\pi \leq 19/2^+$, the model  correctly describes the level 
energies. At higher spins, the experimental level energies are overestimated by 
the model calculations. This is not surprising, since the positive-parity bands 
in the odd-mass, even-$Z$ nuclei from this mass region exhibit a back-bending 
due to a $\nu h_{11/2}$ pair breaking, which is outside the RTRPM model 
space. The model fails in describing the hindrance of the isomeric $M1$ 
transitions to the ground state also, which is attributed to the structure of 
the final state.  These features show the complexity of the low-energy part of 
the $^{105}$Ru spectrum, where single-particle orbits, three-particle clusters 
and collectivity compete.

\section{Acknowledgments}
This work is supported by the Bulgarian National Science fund under contract 
number DMU02/1 and the German BMBF under Contract No. 06 BN 109.

\end{document}